\documentclass[prl,twocolumn,showpacs,preprintnumbers,amsmath,amssymb,superscriptaddress]{revtex4}
\usepackage{graphicx}% Include figure files
\usepackage{dcolumn}% Align table columns on decimal point
\usepackage{color}

\allowdisplaybreaks
\newcommand{\beq}{\begin{equation}}
\newcommand{\eeq}{\end{equation}}
\newcommand{\beqa}{\begin{eqnarray}}
\newcommand{\eeqa}{\end{eqnarray}}

\begin{document}
\title{Dichotomy between large local and small ordered magnetic moment in Iron-based superconductors}

\author{P.~Hansmann}
\affiliation{Institut for Solid State Physics, Vienna University of Technology 1040 Vienna, Austria}

\author{R.~Arita}
\affiliation{Department of Applied Physics, University of Tokyo, Tokyo 153-0064, Japan\\
JST, TRIP, Sanbancho, Chiyoda, Tokyo 102-0075, Japan\\
JST, CREST, Hongo, Tokyo 113-8656, Japan CREST}

\author{A.~Toschi}
\affiliation{Institut for Solid State Physics, Vienna University of Technology 1040 Vienna, Austria}

\author{S.~Sakai}
\affiliation{Institut for Solid State Physics, Vienna University of Technology 1040 Vienna, Austria}

\author{G.~Sangiovanni}
\affiliation{Institut for Solid State Physics, Vienna University of Technology 1040 Vienna, Austria}

\author{K.~Held}
\affiliation{Institut for Solid State Physics, Vienna University of Technology 1040 Vienna, Austria}


\begin{abstract}
We study a four band model for iron-based superconductors 
within
local density approximation + dynamical mean field theory (LDA+DMFT).
This
successfully reproduces the results of models which take As $p$ degrees of freedom explicitly into account and has several \emph{physical} advantages over the standard five $d$-band model.
Our findings reveal that the new superconductors are more strongly correlated than their single-particle properties suggest.
Two-particle correlation functions unveil the dichotomy between local and ordered magnetic moments in these systems, calling for further experiments to better resolve the short time scale spin dynamics.
% needed to omit the following since LDA+DMFT cannot be abbreviated
% , as neutron scattering, femtosecond optics and X-ray absorption.
\end{abstract}
\date{\today}
\maketitle

In the recently discovered iron-based superconductors \cite{kamihara08} the role of electronic correlation is still highly unclear.
Strongly correlated materials are characterized by the presence of large local magnetic moments, which typically long-range order if temperatures are sufficiently low. 
In the proximity of such magnetic phases, also
superconductivity is often observed, with the close-by
magnetic fluctuations  usually cited as evidence for
unconventional (not  phonon-mediated)  superconductivity.
If the local magnetic moment is small and the system is metallic, weak-coupling theories like local spin density approximation (LSDA) \cite{jones89} can be applied.
On the other hand, if the local magnetic moment is large and the exchange coupling between neighboring spins is the dominant interaction, not only the electronic states around the Fermi level but also the higher energy excitations (on the scale of the local Hubbard interaction $U$) are expected to play a role in the superconducting pairing mechanism.
For instance, the latter is definitely the case for cuprates, which are Mott insulating in the absence of carrier doping. 
Iron pnictides instead are metallic and 
undergo  a spin-density wave (SDW) transition below $T\!\approx\!150$K whose characteristics are still under debate.
Understanding the nature of the magnetic properties can therefore also 
help to clarify the origin of superconductivity in these materials.

Experimentally it has been clarified that the different members of the pnictide family have quite different ordered magnetic moments \cite{ishida09}, ranging from 0.3$\mu_B$ (or  0.6$\mu_B$ \cite{Braden10}) in LaFeAsO (1111 compound) to 2.2$\mu_B$ in FeTe (11 compound).
The band structures of these compounds however
do not show distinctive differences, and indeed, LSDA always yields an ordered moment of $\sim$2.0$\mu_B$, for the experimental crystal structures \cite{mazin08,Subedi08a}. 
The failure of density functional theory (DFT) in capturing the correct ground-state properties of pnictides is hence a very important point which needs to be carefully analyzed in order to understand the physics of these compounds:
While usually LSDA underestimates the size of the ordered moment, here the opposite happens.
Even more important than the size of the ordered moment is the fact that
in these systems the magnetic properties are extremely sensitive to the choice of the exchange-correlation functional or of the crystal structure \cite{mazin08}.
In this situation, and also
%% I think we should not indicate  LDA to describe QCP
with the small magnetic moment indicating
the proximity to a quantum-critical point, 
quantum fluctuations 
strongly influence the physics and the moments of iron-based superconductors.
It is therefore natural to conclude that dynamical quantum  fluctuations, not included in LSDA, are crucial for these systems. In particular they can explain the presence of large local magnetic moments which form because of local Coulomb  and exchange interaction 
but only give rise to a much smaller ordered moment at lower temperatures.

There have been many attempts to go beyond LSDA taking electronic correlations more accurately into account. Among these, dynamical mean field theory (DMFT) is one of the most promising, particularly when combined with \emph{ab-initio} band structure calculations \cite{LDADMFTreview}. 
However, the results of such LDA+DMFT calculations for iron-based superconductors \cite{haule08,craco08,anisimov09,aichhorn09,skornyakov09,ishida10} strongly depend on which orbitals are included in the \emph{ab-initio} one-particle Hamiltonian and on the values of the interaction parameters used. 
As a consequence, DMFT calculations have  been employed by different groups in fairly different ways, namely to support that iron pnictides are strongly, intermediately, or  weakly correlated, respectively.
The majority of these studies focused on single-particle spectra and on the comparison with photoemission experiments, except for Refs.~\onlinecite{haule09} and \onlinecite{ishida10} where also the spin susceptibility has been calculated.
In this paper we focus on the dynamics of the local magnetic moment, which we argue is a key indicator for understanding the physics of iron-based superconductors. 
In particular, we conclude that in the single-particle spectral function 
correlation effects are hardly visible, while, at the same time, the spin-spin correlation function reveals the existence of a large local magnetic moment.
This turns out to be crucial for the explanation of some controversial experimental results in these systems.

For iron-based superconductors, two classes of models have been proposed \cite{vildosola08,miyake08}: 
One is a $d$ only model which takes  the Fe 3$d$ degrees of freedom into account, while the others are  $dp$ or $dpp$ models  considering pnictogen/chalcogen $p$ and O 2$p$ electrons explicitly. 
In this work, we take the $d$ model as a starting point.
The $d$ models considered hitherto, however, pose
some physical and technical problems:
Each Wannier function of $d$ character has a fairly different spread in real space, due to the orbital-dependent hybridization between $p$ and $d$ \cite{vildosola08}. 
Thus the interaction parameters strongly depend on the orbital \cite{nakamura08,miyake10}. 
In our coordinate system, in which $x$- and $y$-axes point to the pnictogen/chalcogen atom, the $3z^2$-$r^2$ orbital, for example, has a small $p$-$d$ hybridization, so that it is well localized. 
On the other hand, the $x^2$-$y^2$ orbital has long tails in the direction of the pnictogen or chalcogen sites. 
Such an orbital dependence causes problems when calculating the self-energy due to electron correlation: 
In order to avoid a 
double-counting of correlation effects already considered within LDA, one would have to introduce an \emph{ad hoc} orbital-dependent 
level shift in the many-body calculation. 
This level shift has been realized to be particularly important for the $3z^2$-$r^2$ orbital. 
In calculations based on the so-called FLEX approximation \cite{ikeda08}, it has been shown that the $3z^2$-$r^2$ level becomes higher in energy and makes a large Fermi surface not present in LDA. 
A similar tendency is also seen in DMFT calculations \cite{ishida10}, namely the $3z^2$-$r^2$ occupancy gets dramatically smaller than in LDA depending on the
strength of interaction parameters, contrary to what happens in LDA+DMFT calculations for $dp$ and $dpp$ models \cite{anisimov09,skornyakov09,aichhorn09}. 
In order to overcome this problem, some authors added a constant part to the self-energy \cite{ikeda08}, or constrained the zero frequency value of the self-energy to get the appropriate orbital shift \cite{ikeda10}. 

\begin{figure}[ht]
\begin{center}
\includegraphics[width=7.5cm]{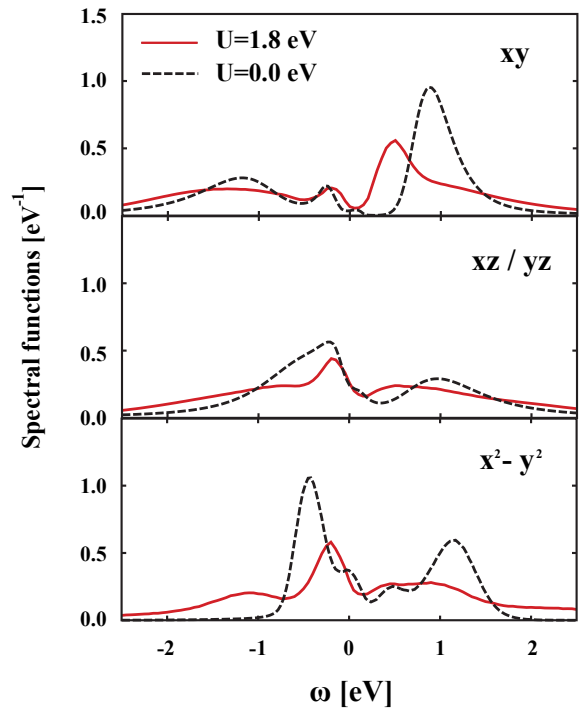}
%{4x4J045spec.new.eps}
%{{\resizebox{8.0cm}{!}{\includegraphics {4x4J045spec.new.eps}}}}
%{\rotatebox{-90}{\resizebox{6.0cm}{!}{\includegraphics {fwdspecsmoment/fermispecs.ps}}}}
\end{center}
\vspace{-0.5cm}

\caption{Orbital-resolved spectral function of the four-band model for LaFeAsO at $T\!=\!460$K, for  $U\!=\!1.8$eV,$J\!=\!0.45$eV (solid lines) and $U\!=\!J\!=\!0$ (dashed lines). Including electronic correlations does not change the spectrum drastically.}
\label{spectra}
\end{figure}

Here we take another route and assume the $3z^2$-$r^2$ to be fully occupied. We therefore do not include it in our low-energy Hamiltonian for LaFeAsO and are left with a model with merely four equally correlated 
orbitals \cite{arita09}, which allows us to circumvent an orbital
dependent double-counting adjustment. 
This approximation is justified by the fact that the band with mainly $3z^2$-$r^2$ character lies below the Fermi level.
We argue that the results of the four orbital model compare to the $dpp$ model much better than the five band one. 
It has been already shown \cite{arita09} that the Fermi surface of the four band model is almost exactly the same as that of the five band model in LDA,
and does not change so drastically even after the inclusion of many-body effects.

The four and the five $d$-band models have of course different values of the interaction parameters. 
Constrained random phase approximation (RPA) calculations for the five band model give an intra-orbital Coulomb interaction $U$ of about $2.2-3.3$eV and a Hund's rule coupling of about $0.3-0.6$eV 
\cite{nakamura08}. In our case we have to consider the screening effect of $3z^2$-$r^2$ orbital as well as the slightly more extended tails of the Wannier functions which both reduce the value of $U$. For this reason, hereafter we use $U\!=\!1.8$eV. Numerical limitations necessitate a Hund's exchange $J$ of
Ising-type for four orbitals. Below, we will show however
a comparison of results between  Ising and  full SU(2) symmetric interaction, for a model with fewer orbitals. The differences turn out not to be relevant for the present discussion.

In Fig.~\ref{spectra}, we show the spectral functions of the four band model for non-interacting electrons (dashed lines) and for $U\!=\!1.8$eV, $J\!=\!0.45$eV and temperature $T\!=\!460K$ ($\beta\!=\!25\,$eV$^{-1}$).
%None of the bands shows dramatic changes with correlation up to $\approx -0.5$eV.
Including $U$ and $J$ renormalizes (shrinks the width of) the minimum 
structure around the Fermi level. At the same time, some of the spectral weight is shifted to Hubbard-like shoulders at higher energies, in a way such 
that the overall
bandwidth and spectrum remains close to the non-interacting one.
The values of the quasiparticle weights are 
$Z\sim\!0.51$, 0.45, and 0.60  for $yz$($xz$), $x^2$-$y^2$, and 
$xy$ band, respectively, in agreement with %  angular resolved
photoemission spectroscopy (PES) \cite{Malaeb08,Qazilbash09}.
As anticipated above, the results of the four orbital model resembles very closely the $dp$ and the $dpp$ models \cite{skornyakov09,aichhorn09}.
For instance, in our $xy$ band the peak around $\sim\!1$eV shifts towards lower energies, very similarly to the corresponding spectral function in Ref. \onlinecite{skornyakov09}  and  Ref. \onlinecite{aichhorn09} (there denoted as $x^2$-$y^2$ orbital as their coordinate system is rotated by 45$^\circ$).
On the other hand, our $x^2$-$y^2$ has a structure around $\sim\!1$eV coming form the hybridization to the $3z^2$-$r^2$ and the pnictogen/chalcogen $p$, which are included effectively in our four band Hamiltonian.
We can therefore conclude that the four band model reliably reproduces the results of photoemission experiments with the clear advantage 
% Ale, Philipp I did not include your suggestion in comparison with 5band
% since I think it is also w.r.t. dpp
of (i) having a smaller number of parameters and (ii) of yielding a set of $d$ orbitals with much more similar spatial spread (and double counting correction).

The picture arising from merely  analyzing the spectral function hence suggests that iron-based superconductors are quite far from being standard strongly correlated materials, such as cuprates or other transition-metal compounds.
On the other hand, calculations based on FLEX which one would expect to work for weakly correlated materials, here fail to reproduce the correct stripe pattern of antiferromagnetic spin fluctuations \cite{arita09}. 
This indicates that  iron-based superconductors cannot be categorized 
as weakly correlated systems, at least not with respect to their 
 two-particle correlation functions.

These considerations naturally leads to the question: Are iron-based superconductors more correlated than
 their single particle quantities such as the
 photoemission spectra suggest? 
To answer this question we calculated the local spin susceptibility of LaFeAsO within LDA+DMFT and study whether or not this indicates the existence of a large local magnetic moment in these compounds.

\begin{figure}[tb]
\begin{center}
%{{\resizebox{8.0cm}{!}{\includegraphics {fwdspecsmoment/new_normalized_compareb10_b25.ps}}}}
\includegraphics[width=8cm]{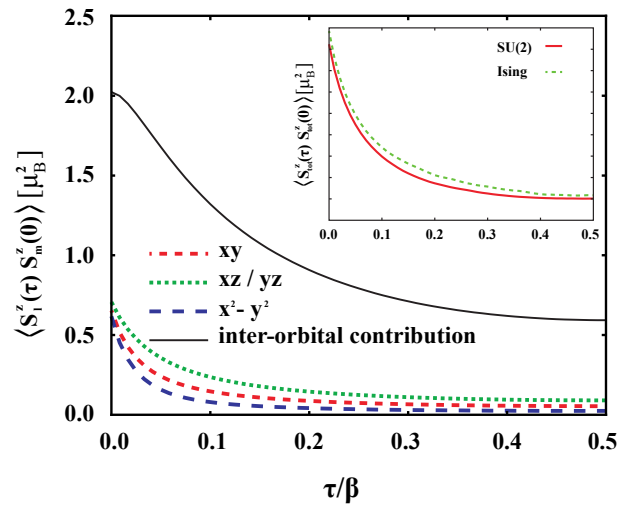}
%{spincorrJxb25.eps}
\end{center}
\vspace{-.5cm}

\caption{Spin-spin correlation function  for $J\!=\!0.45$eV and $\beta\!=\!25$eV$^{-1}$. We plot the different intra-orbital contributions and the sum of all inter-orbital contribution.
This orbital-resolved 
presentation clearly shows that the $J$-induced inter-orbital correlation 
is particularly large. Inset: 
 Comparison between Ising- and SU(2)-symmetric Hund's exchange for a related two-band model, showing only quite small differences.}
\label{SS_notnorm}
\end{figure}

In Fig.~\ref{SS_notnorm}, we plot the
(dynamical)  local spin-spin correlation function  
$\chi_{lm}(\tau)  \equiv \langle S^z_l(\tau) S^z_m(0) \rangle$ 
for (imaginary) time $\tau$. 
Resolved are its  intra-orbital ($l=m$) and inter-orbital ($\sum_{l\neq m}  \chi_{lm}(\tau) $) contribution.
Similar to the case of one-particle properties,
the $U$-driven intra-orbital spin correlation
 is only slightly enhanced in comparison to the non-interacting value
which for equal times ($\tau=0$) is $0.5 \mu_B^2$
\cite{footnotenonint}. In stark contrast, the 
inter-orbital contribution
 which  vanishes without interaction
is strongly enhanced. This reflects the strong tendency of the system to align spins between different orbitals. It can be understood by noting that, since the crystal field splitting is small ($\sim\!0.2$eV) \cite{arita09}, even an intermediate value of the Hund's rule exchange $J$ is very effective. 
Hence, the large inter-orbital spin correlation function
points to the importance $J$ plays for inducing 
electronic correlations in iron-based superconductors.

Let us note in passing the inset  of Fig.~\ref{SS_notnorm}
which compares the total $\chi(\tau)$ for a half-filled two-band Hubbard model with a semi-elliptic density of states with bandwidth 4eV, $U\!=\!2.5$eV, $J\!=\!0.5$eV and $\beta\!=\!20$eV$^{-1}$, a choice that gives  renormalization factors very similar to our four-band realistic calculation. 
As  already mentioned, the close agreement of the two curves allows us to exclude that the Ising approximation for the $J$ term has a big influence in the relevant parameter regime.

\begin{figure}[tb]
\begin{center}
%{{\resizebox{8.0cm}{!}{\includegraphics {fwdspecsmoment/new_normalized_compareb10_b25.ps}}}}
\includegraphics[width=8cm]{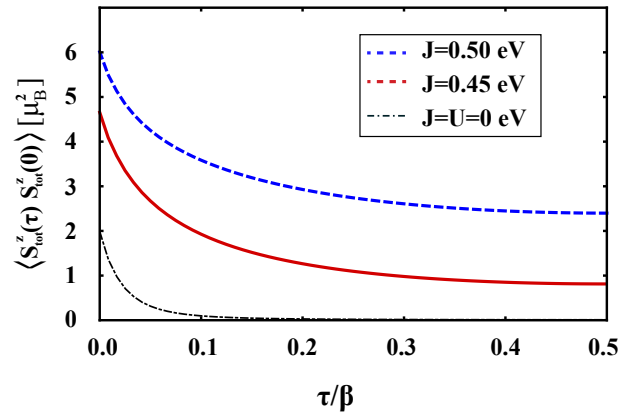}
\end{center}
\vspace{-.5cm}

\caption{Total  spin-spin correlation function for LaFeAsO at two different values of $J$ and $\beta\!=\!25\,$eV$^{-1}$, compared  to
the non-interacting  $U\!=\!J\!=\!0$ case. 
 The short-time ($\tau\!=\!0$) 
LDA+DMFT local moment  for $J\!=\!0.45$eV is
$m_{\rm loc}\!=\! 2.16 \mu_B$, comparably large as in LSDA;
whereas the long-time
moment is screened to only $m\!\approx\!0.7 \mu_B$ at $T\!=\!50\,$K, see text.}
\label{SS_norm}
\end{figure}

Turning to the  central result of our paper,
the total  $\chi(\tau)$  in  Fig.~\ref{SS_norm},
we see a much larger susceptibility  than in the non-interacting case.  
This enhancement of $\chi(\tau)$  is a measure for the correlations
in  iron-based superconductors.
At $\tau\!=\!0$, the susceptibility $\chi(\tau)$
 gives us a direct measure of the bare local moment
$m_{\rm loc}\!=\!\sqrt{\chi(0)}$. This local moment corresponds to
the responses on short time  (or high energy) scales and is quite
large, i.e.,
$m_{\rm loc}\!=\! 2.16 \mu_B$ and $2.45 \mu_B$ for $J\!=\!0.45\,$eV and $0.5\,$eV, respectively.
Note that such  variations of $J$ are realistic when going from
LaFeAsO to FeTe. 
While the  $J\!=\!0.45\,$eV value of $m_{\rm loc}$ is only slightly $T$-dependent
($m_{\rm loc}\!=\! 2.36 \mu_B$ at  $\beta\!=\!10\,$eV$^{-1}$), at $J\!=\!0.5\,$eV an even larger moment is formed
at  higher temperature
($m_{\rm loc}\!=\! 3.44 \mu_B$ at  $\beta\!=\!10\,$eV$^{-1}$).

While this local moment is large and indicates strong correlations,
it is not the one which was hitherto measured experimentally.
Experiments such as magnetic susceptibility measurements,
nuclear magnetic resonance, M{\"o}{\ss}bauer spectroscopy and muon relaxation are slow compared to the electronic dynamics on the $fs$ time scale.
Hence, these experiments correspond to larger $\tau$'s or the integrated
(static) susceptibility
$\chi(\omega=0)=\int_0^{\beta} {\rm d} \tau \chi(\tau)$.
A central result of our calculation is that this long-time susceptibility 
or
a corresponding 
magnetic moment, which one can define
by resolving $\chi(\omega=0)=m^2/T$ for $m$,
is strongly reduced (screened) compared to the instantaneous
local magnetic moment $m_{\rm loc}$.
 Already at $\beta=25\,$eV$^{-1}$ ($\beta=10\,$eV$^{-1}$)
the dynamic screening leads to
strongly reduced moments of
$m= 1.2 (1.9)$ and 1.8(3.4) for $J=0.45\,$eV and $0.5\,$eV, respectively.
And for lower temperatures  these values are 
much further reduced
because of  the screening.
At $T=50\,$K, i.e., in the temperature range where
such magnetic moments were experimentally measured,
 an extrapolation of our data
yields a crude (overestimated) approximation of $m \approx 0.7 \mu_B$ \cite{noteextrapol,noteJ1J2}.
Since also
the SDW phase of iron-based superconductors
is very itinerant for both spin species,
 we expect that
the dynamical screening identified here as the origin
of the smallness of the  (long-time)  magnetic moment 
   survive (to a large extent) also in the magnetic phase.

In conclusion, LDA+DMFT predicts that the local magnetic moment in iron-based superconductor is, in the paramagnetic phase, comparable to  the
ordered moment of LSDA. This moment is formed due to 
a local Hund's rule spin alignment.
However, there is
a  dichotomy between this local magnetic moment and the dynamically
screened moment, which is much smaller and beyond LSDA.
The latter 
is actually much smaller in LDA+DMFT.
 Experiments performed
hitherto measured  the low-energy (or long-time) moment, i.e.,
the dynamically screened one. For measuring the (bare) local moment,
experimental measurements on the time scale of $fs$ are needed.
A possibility to this end is to integrate 
 neutron scattering measurements over ${\bf Q}$ and $\omega$.
For such an experiment, our calculations predict
an intermediate-to-large value of the local magnetic moment.
Similarly, X-ray spectroscopy is a very promising technique for measuring the size of the local magnetic moment, but so far this has been mainly used to estimate the strength of the interaction parameter, like, e.g., in Ref. \onlinecite{yang09}.
Such experiments, if performed, will clarify whether our idea of iron-based superconductors being more strongly correlated than what is naively  expected from photoemission experiments, is correct. 
This can eventually settle the role electronic correlations play 
in the new class of iron-based superconductors.

We acknowledge financial support from the EU network MONAMI and the FWF through the ``Lise-Meitner'' grant n.M1136 (GS) and science college WK004 (PH). We thank M.~Aichhorn, O.~K.~Andersen, L.~Boeri, A.~Georges, H.~Ikeda, M.~Imada, K.~Nakamura, T.~Miyake, J.~Kunes, and G.~Kotliar for discussions as well as the KITP Santa Barbara for hospitality.

%Unused bibitems

%\bibitem{delaCruz08}C.\ de la Cruz, Q.\ Huang, J.\ W.\ Lynn,
%J.\ Li, W.\ Ratcliff II, J.\ L.\ Zarestky, H.\ A.\ Mook,
%G.\ F.\ Chen, J.\ L.\ Luo, N.\ L.\ Wang, and P.\ Dai:
%Nature (London) {\bf 453} 899 (2008).

\begin{thebibliography}{99}
\bibitem{footnotenonint}
This stems from the fact  if there is a spin $\sigma$ in orbital $l$
at time 0, looking at time $\tau\!=\!0$ again it is still there.
\end{thebibliography}
\end{document}